\documentclass[usegraphicx,usenatbib]{mn2e}

\begin{document}

\title[Star formation and spin chirality]{Galaxy Zoo: A correlation between coherence of galaxy spin chirality and star formation efficiency\thanks{This publication has been made possible by the participation of more than 100,000 volunteers in the Galaxy Zoo project. Their individual contributions are acknowledged at \texttt{http://www.galaxyzoo.org/Volunteers.aspx}}}

\author[Jimenez et al.]{Raul Jimenez$^{1}$\thanks{email:raulj@astro.princeton.edu}, An\v{z}e Slosar$^{2,3,4}$, Licia Verde$^{1}$, Steven Bamford$^{5}$, Chris Lintott$^{3}$, 
\and Kevin Schawinski$^{7,8}$, Robert Nichol$^{6}$, Dan Andreescu$^{9}$, Kate Land$^{3}$, Phil Murray$^{10}$, 
\and M. Jordan Raddick$^{9}$, Alex Szalay$^{11}$, Daniel Thomas$^{5}$, Jan Vandenberg$^{11}$\\
$^1$ICREA \& Institute of Space Sciences (CSIC-IEEC), Campus UAB, Bellaterra 08193, Spain \\
$^2$Berkeley Center for Cosmological Physics, Lawrence Berkeley Nat. Lab \& Phys. Dept, University of California, Berkeley, CA 94720, USA \\
$^3$Astrophysics Department, University of Oxford, Oxford, OX1 3RH \\
$^4$Faculty of Mathematics \& Physics, University of Ljubljana, Slovenia \\
$^5$Centre for Astronomy and Particle Theory, University of Nottingham, University Park, Nottingham, NG7 2RD, UK \\
$^6$Institute of Cosmology and Gravitation, University of Portsmouth, Mercantile House, Hampshire Terrace, Portsmouth, PO1 2EG, UK \\
$^{7}$Department of Physics, Yale University,  New Haven, CT 06511, USA \\
$^{8}$Yale Center for Astronomy and  Astrophysics, Yale University, P.O. Box 208181, New Haven, CT 06520, USA\\
$^{9}$LinkLab, 4506 Graystone Ave., Bronx, NY 10471, USA \\
$^{10}$Fingerprint Digital Media, 9 Victoria Close, Newtownards, Co. Down, Northern Ireland, BT23 7GY, UK \\
$^{11}$Department of Physics and Astronomy, The Johns Hopkins University, Homewood Campus, Baltimore, MD 21218, USA} 

\maketitle

\begin{abstract}
We report on the finding of a correlation between galaxies' past star formation activity and the degree to which neighbouring galaxies rotation axes are aligned.  This is obtained by cross-correlating star formation histories, derived with MOPED, and spatial coherence of spin direction (chirality), as determined by the Galaxy Zoo project, for a sample of SDSS galaxies.  Our findings suggest that spiral galaxies which formed the majority of their stars early ($z > 2$) tend to display coherent rotation over scales of $\sim 10 \rmn{Mpc}/h$. The correlation is weaker for  galaxies  with significant  recent star formation. We find evidence for this alignment at more than the $5\sigma$ level, but no correlation with other galaxy stellar properties. This finding can be
  explained within the context of hierarchical tidal-torque theory if  the SDSS galaxies harboring the majority of the old stellar
  population where formed in the past, in the same filament  and at about the same time. Galaxies with significant recent star formation instead are in the field,  thus  influenced by the general tidal field that will align them in random directions or had a recent merger which would promote star formation, but deviate the spin direction.
\end{abstract}


\section{Introduction}

In our progress to understand how galaxies have formed and evolved, it
has become recently clear (e.g. \citet{Jimenez08}) that the present
stellar mass of galaxies determines most of the galaxy properties. It
is thus possible to predict \emph{on average} what the star formation,
metallicity, environment, etc. a galaxy has by simply measuring its
current stellar mass. However, second order differences among the
properties of galaxies with the same stellar mass exist and need to be
explained. One of the most obvious physical mechanisms to explain this
second parameter is the spin of the dark matter halo. It is now 60
years since Hoyle's seminal paper on the subject \citep{Hoyle49} that
showed how angular momentum in galaxies can be generated via the tidal
field of the other galaxies. Later on, \citet{doro70} developed the
tidal torque theory within the framework of hierarchical galaxy
formation that determines the amplitude and direction of the spin of a
dark matter halo based on the surrounding dark matter field (see
\citet{Schaefer08} for a recent review). Numerical N-body simulations
produce results that are in good agreement with the theoretical
predictions, although linear theory is not always sufficient to
determine the final angular momentum of a collapsed object (e.g.,
\cite{BarnesEfstathiou87, Porciani02a} and references therein). In
addition mergers are expected to significantly alter a halo final spin
\citep{merger}.

The amplitude of the dark matter halo spin will influence the radius
where baryons will settle into a disk
\citep{FallEfstathiou80,WhiteRees78}, thus influencing its density and
therefore the star formation history of the galaxy. The influence of
spin on star formation history has been studied in detail
\cite{Toomre64, Dalcanton97,JHH97,Mo98, AvilaRese98}.

One interesting feature of tidal-torque theory in hierarchical models
\citep{HeavensPeacock88,CatelanPorciani01,Catelanlensing,Crittenden2001,Porciani02a,HahnSpins07}
is the prediction of correlated spin directions and that the spin
direction for dark halos is strongly influenced by the halo
environment. \citet{Pen2000} reported a detection of galaxy spin correlations at
97\% confidence and \citet{SlosarSpins} have measured the correlation
function of the spin chirality and report, for the first time, of a
signal at scales $< 0.5$ Mpc/$h$ at the $2-3 \sigma$ level. This
subject has received renewed interest not just for being a test of
tidal torque theory but because the mechanism that produces angular
momentum alignment is believed to create correlations between observed
galaxy shapes, introducing a potential contamination to the
cosmological weak lensing signal.

Observationally, the full information on the spin vector of a dark
matter halo is very difficult to obtain. Concentrating on disk
galaxies, the plane of the disk determines the axis of rotation of the
disk of baryonic matter. For galaxies seen in projection, the observed
galaxy ellipticity constrains the galaxy spin axis. For infinitely
thin and perfectly circular disks, the spin vector axis would be known
but not the spin "chirality" (the spin direction along the axis). For
realistic disks, projection effects mean that what can be measured
reliably is the projected spin axis.  Information abut the spin
chirality is absent in the study of galaxies ellipticity. Nevertheless, the chiral
information is known for the sample of face-on galaxies of the Galaxy
Zoo project \footnote{{\tt www.galaxyzoo.org}}. Correlation of
chirality therefore implies a correlation of the spin vectors.

An expectation of tidal torque theory is that haloes which formed together, and thus
experienced a similar tidal field during their initial collapse, will have similarly aligned 
spin vectors. Galaxy chirality should thus be coherent on scales related to those of large scale structure.

Here we explore whether spatially coherent spin chirality (and therefore spin
vectors) correlate with other galaxy properties that
depend on the galaxy stellar population and star formation
history. Our main finding is that the {\em absolute} value of the average spin
direction of the galaxies located in  a spatial patch (pixel), is correlated with past star formation activity in these galaxies, while we find no correlation with other galaxy properties like metallicity, or with the average spin itself, i.e. the
universe does not have a preferred direction.
The rest of the paper is organized as follows: \S 2 describes the
galaxy sample selection and the methodology, \S 3 presents the
results. In \S 4 we present some discussion and draw our conclusions.

\section{Sample selection and method}

In the Galaxy Zoo project, a sample of 893,212 galaxies were visually
classified by about 90,000 users. The sample was selected to be
sources that were targeted for SDSS spectroscopy, that is extended
sources with Petrosian magnitude $r < 17.77$. Additionally, we
included objects that were not originally targeted as such, but were
observed to be galaxies once their spectrum was taken. Where
spectroscopic redshifts were available, we found that  galaxies have the mean
redshift of $z=0.14$ and the objects with the highest redshift reach
$z\sim 0.5$.  The galaxies thus probe our local universe at
cosmological scales. Each object has been classified about $40$ times
from a simplified scheme of 6 possible classifications: an elliptical,
a clockwise spiral galaxy, an anti-clockwise spiral galaxy, an edge-on
spiral galaxy, a star / unknown object, a merger.  Various cuts
(hacking attempts, browser misconfigurations, etc.)  removed about 5\%
of our data. The data were reduced into two final catalogues based on
whether data was weighted or unweighted. In the unweighted data, each
user's classification carried an equal weight, while in the weighted
case, users weights were iteratively adjusted according to how well
each user agreed with the classifications of other users. In both
cases, the accrued classifications were further distilled into
``super-clean", ``clean" and ``cleanish" catalogs of objects, for which we
required 95\%, 80\% and 60\% of users to agree on a given
classification.  In all cases, this is a statistically significant
classification with respect to random voting; however, the human
``systematical'' error associated with it is difficult to judge.  In
any case, we are in the limit where taking more data will not change
our sample beyond noise fluctuations as the votes are uncorrelated.  A
detailed account of the reduction procedures as well as the procedure
to measure spin orientations, and removal of systematics is explained
in  \cite{lintott08} and \cite{Land08}.

We consider the sample of spiral galaxies of the Galaxy Zoo
project for which the spin chirality (i.e. the direction of the galaxy arms
winding) has been determined.  This piece of information for each
galaxy is what we refer to as halo spin chirality.  In general, the
gas of a galaxy should be rotating in the same direction as the halo:
about 4\% of the galaxies do not obey this \citep{Pasha82}, but there
should be a strong --although not perfect--correlation between the
angular momentum vector of gas and that of the dark matter halo
hosting a galaxy \citep{vandenBosch02}. Here we assume that the
chirality of the galaxy is a proxy for the rotation direction of the
host dark matter halo.

Our catalog of galaxy star formation properties is obtained from
\citet{Panter07}. This catalog has been constructed from the
spectroscopic main galaxy sample of the SDSSS-DR4 data release
\citep{dr4}. Star formation histories, metallicities, stellar masses
and dust content, have been extracted using the MOPED \citep{HJL00}
algorithm, which allows for rapid extraction and parameter exploration
of galaxy spectra using stellar population models. The method has been
explored in detail in \citet{Panter07} and has been tested with the
VESPA \citep{rita} algorithm for accuracy and extent of degeneracies
in the recovered parameters (see \cite{Panter07,rita} for details). 
The star formation histories catalog from VESPA\citep{vespa} 
is available online\footnote{{\tt http://www-wfau.roe.ac.uk/vespa/}}.
While for the SDSS spectra, individual star formation histories are
poorly constraint \citep{rita} because of the low $S/N$, average
properties sample are robust and well constrained; it is these average
quantities that we use in our study.  \citet{Panter07} give a thorough
and detailed explanation of the method, its shortcomings and
advantages and we refer the interested reader to this paper for a
detailed description of the star formation catalog used in the present
work. The catalog contains about half million galaxies.

\begin{figure}
\includegraphics[width=1.1\columnwidth,angle=0]{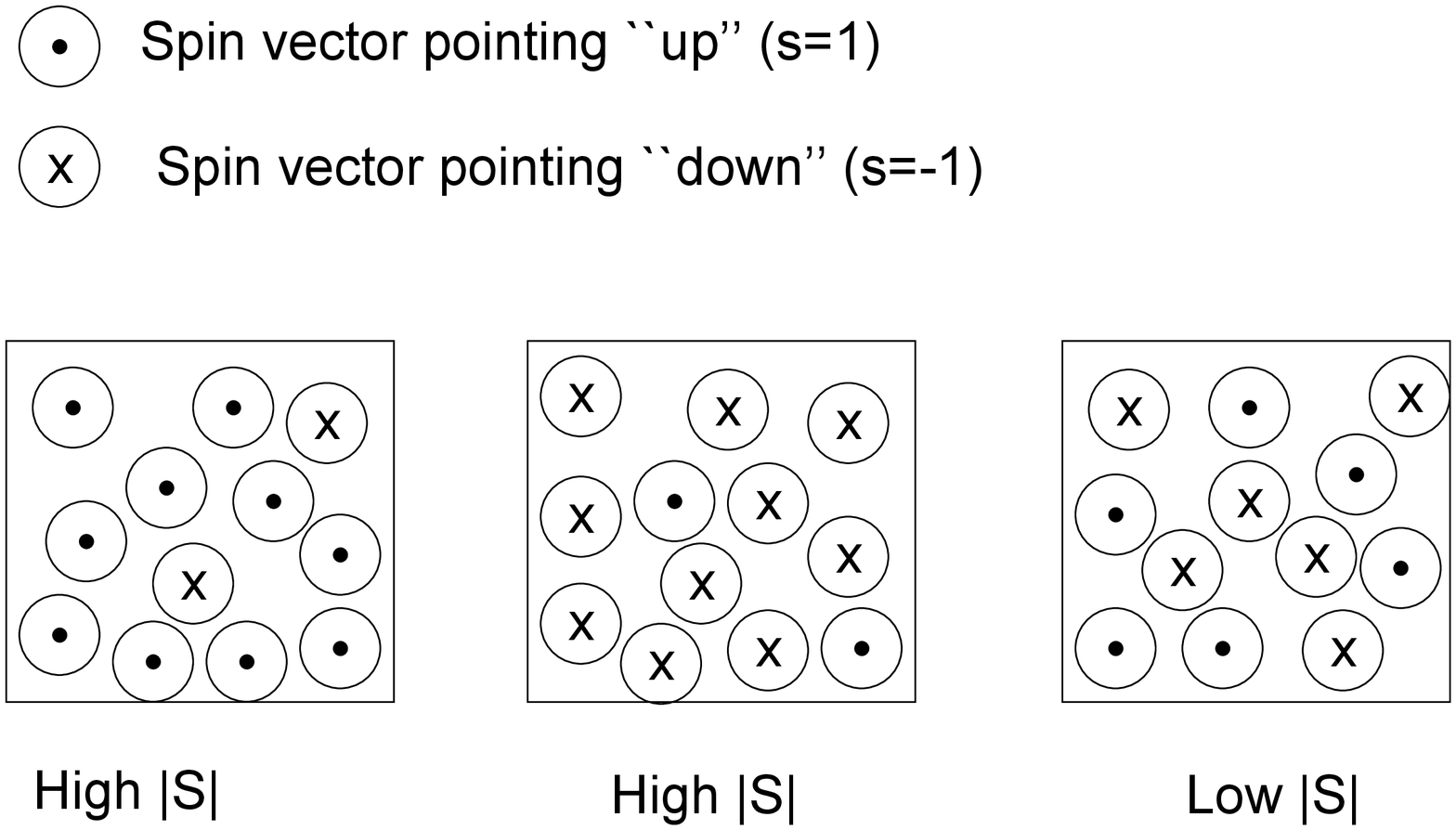}
\caption{\label{fig:spincoherence}
Graphical depiction for a given pixel size of what it means to have high value of $s_j$.}
\end{figure}

To create our final sample we match the above two catalogs, MOPED and
Galaxy Zoo (keeping only galaxies that are classified as clean), to obtain a total of $12897$ galaxies at redshift
$z<~0.2$.  The redshift distribution of our sample is fairly narrow
(the redshift interval $0.012<z<0.13$ encloses 90\% of the galaxies,
the maximum of the redshift distribution is $z_{m}=0.08$. It is this sample that we use in our study.
Note that the galaxies in the sample are spirals, and therefore the downsizing effect is not as extreme as in ellipticals.

We uniformly pixelize the survey using a range of pixel sizes. In what follows we
use a flat concordance LCDM model to convert between redshifts and
distances.  In each pixel we define a pixel spin chirality $S_j$ given
by the average spin chirality for the galaxies in the pixel:
$S_j=\sum_{i=1}^{N_j} s_i/N_j$ where $N_j$ denotes the number of
galaxies in pixel $j$ and $s_i$ denotes a galaxy spin chirality; $s_i$
can only take values of $+1$ or $-1$ while $-1<S_j<1$. For
uncorrelated spin direction, within the errors $S_j=0$, only for
correlated spin direction the average spin chirality will be
significantly non-zero (see Fig.~\ref{fig:spincoherence}). Note that when we compute the
cross-correlation of galaxies quantities with the pixel spin chirality  $S_j$ we  will use the absolute value, $|S_j|$. We will show that there
is no correlation when we use the full value  of $S_j$ (including its sign).

Each galaxy in the sample (index $i$) has associated a stellar mass, a
metallicity, a metallicity history and a star formation history. The
star formation history is described by the star formation in
independent bins (index $\beta$), $\psi_{\beta,i}$: 9 equally spaced
in look back time and two bins for high redshift star formation (see
table 1).  For each galaxy, the star formation is normalized to unity.
For each pixel we also define a deviation from the mean star formation
history given by the average star formation in each bin over the pixel
galaxies minus the global average star formation:
$SF_{j,\beta}=\sum_{i=1}^{N_j} \psi_{\beta,
  i}/N_j-\langle\psi\rangle$. Metalicity history is treated in the
same way.
Note that by comparing average quantities in each pixel rather than total quantities we are not directly sensitive to the galaxy local density.  Correlation between chirality and density has been explored in 
\cite{Land08}.

We therefore have several pixelized maps: a map of spin chirality, and
24 maps of galaxy star formation properties (stellar mass,
metallicity, 11 maps of star formation at different look back times,
and similarly 11 maps for metalicity history).

To quantify a possible correlation between galaxy properties and spin
coherence we use the Pearson correlation coefficient and explore how
it varies as a function of pixel size, lag, and star formation
properties.  Errors are estimated by repeating the procedure with
randomized spins. This cross-correlation in pixels of a finite size
essentially contains information on integrated cross-correlation
function below some pixel size. If certain properties correlate in
two-point statistics up to a certain scale, our statistical test will
pick it up.

\section{Results}

\begin{table}
\caption{Centre and boundaries ((l)ower, (c)enter, (u)pper) of the star formation  bins in redshift ($z$) and
look-back time $t_{\rm lb}$ in Gyr} \label{table:bins}
\begin{tabular}{l||c|c|c||c|c|c}
\hline
bin  & l $z$&c $z$&u $z$&l $t_{\rm lb}$&c $t_{\rm lb}$&u $t_{\rm lb}$ \\
\hline
11   & 0.0007  &  0.001 & 0.00145 &  0.00966        & 0.014   &  0.0200    \\
10   & 0.00145 & 0.0021 & 0.003   &  0.0200         & 0.029   &  0.0414     \\
9   & 0.003   &  0.006 & 0.0063  &  0.0414         & 0.06    &  0.0857      \\
8   & 0.0063  &  0.012 & 0.013   &  0.0857         & 0.12    &  0.1776      \\
7   & 0.013    & 0.0179 & 0.027   &  0.1776         & 0.26    &  0.3677      \\
6   & 0.027    & 0.0419 & 0.057   &  0.3677         & 0.53    &  0.7614      \\
5   & 0.057    & 0.0839 & 0.125   &  0.7614         & 1.10    &  1.5767      \\
4   & 0.125   & 0.186  & 0.287   &  1.5767         & 2.27    &  3.2650     \\
3   & 0.287   & 0.456  & 0.786   &  3.2650         & 4.70    &  6.7609      \\
2   & 0.786   & 1.200  & 2.000   &  6.7609         &   8.46    &  10.32    \\ 
1  & 2.000   & 4.000  & 8.000   &  10.32         &   12.09   &   13.00     \\
\hline
\end{tabular}
\end{table}

\begin{figure}
\includegraphics[width=\columnwidth,angle=0]{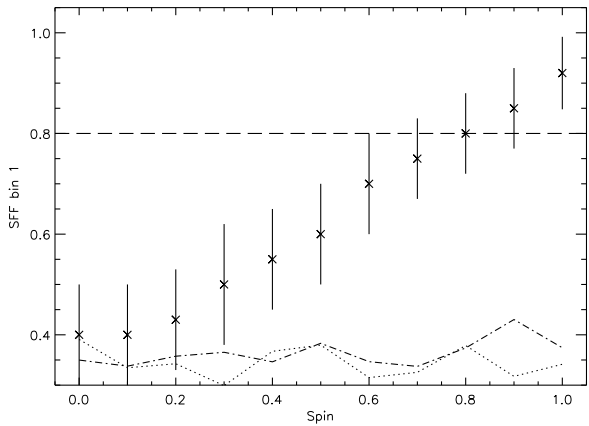}
\caption{\label{fig:chiralold}
Relation between pixel-averaged star formation fraction  in the oldest MOPED bin (1, see Table~1) and pixel-averaged  absolute spin $|S_j|$ (for a pixel size of $10 Mpc/h$) for all those galaxies that have formed more than 30\% of their present stellar mass at $z > 2.0$. Due to the large number of pixels,  instead of showing a scatter plot we sow the mean relation, error-bars show the standard deviation around the mean relation.
Note that spiral galaxies that formed more than 30\% of their stars at $z > 2$ have preferentially a coherent spin chirality. If we focus on those spiral galaxies that formed more than 80\%, denoted by the dashed line, of their stars at $z > 2$ then virtually {\em all} of them have coherent spin chirality. The dotted and dash-dotted lines show the pixel-averaged star formation but for choices of the pixel size of $2 Mpc/h$ and $50 Mpc/h$ respectively. For these cases the value of the star-formation fraction as a function of absolute spin is consistent with zero as error bars are similar to the $10 Mpc/h$ pixel-size case and are not plotted for clarity (see text for more details).}
\end{figure}

\begin{figure}
\includegraphics[width=\columnwidth,angle=0]{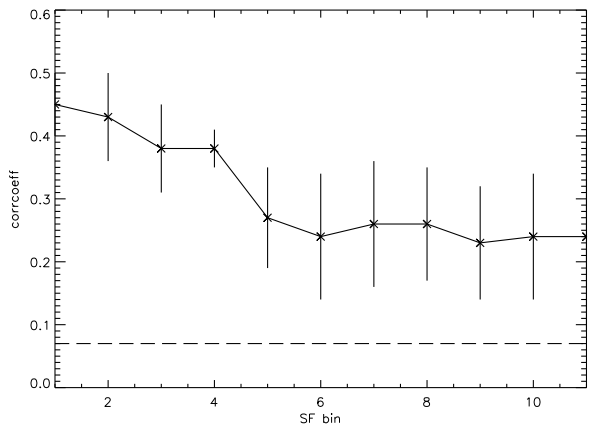}
\caption{\label{fig:corrspins} Pearson correlation coefficient  between absolute value of  pixel spin chirality and past star formation  as a function of the start formation in look-back time bin number (see Table~1 for conversion between bin number and look-back time). The pixel size is $10$Mpc/$h$. The points are the mean of  the correlation coefficient for several redshift slices. Error bars  show  jacknife estimates of the correlation coefficient {\it rms} by using several redshift slices, the dashed line shows the  $1-\sigma$ correlation level obtained with a randomized spin map. Galaxies with star formation in the oldest bin (1-3)  show strong correlation with spin chirality (and therefore with spin alignment). The correlation decreases  for galaxies with more recent star formation.}
\end{figure}

We are interested in exploring if galaxies with a particular property
have their spins aligned or randomly oriented with respect to their neighbours. In this respect we do
not need to worry about projecting the spin vector on a common
reference frame for all galaxies as was done by
\citet{Land08,SlosarSpins} as we are only interested in the
cross-correlation between the spatially averaged chirality of the galaxies and their star
formation properties.  We also interpret the chirality of the galaxy
spin as a proxy for the halo spin chirality.

Spins of the galaxies in our sample are spatially correlated
\citep{SlosarSpins} and star formation properties are also spatially
correlated with the large-scale cosmological structures (e.g.,
\cite{marks}). The cross correlation as defined in \S 2 will indicate which galaxy
property is mostly correlated with the large-scale tidal field
responsible for spin alignment.

\begin{figure}
\includegraphics[width=\columnwidth,angle=0]{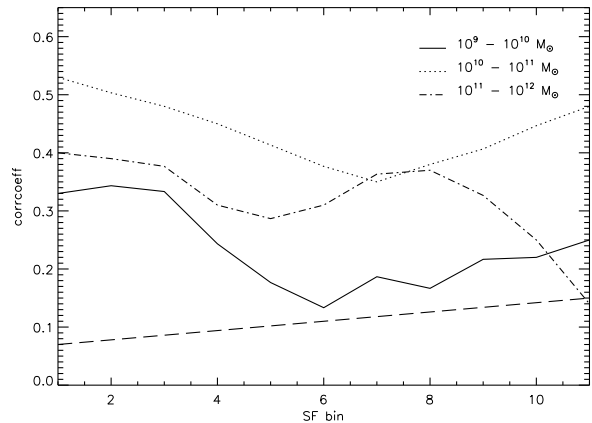}
\caption{\label{fig:corrmass} Same as Fig.~\ref{fig:corrspins} but for three different mass ranges. In this cases we use the projected map in order to increase the signal-to-noise. Note that the signal at bins $1-2$ is dominated by galaxies in the mass range $10^{10}-10^{11}$ M$_{\odot}$. }
\end{figure}

We find no correlation of the absolute value of the pixel spin chirality with stellar
mass, metallicity or metallicity history but we find significant
correlation with star formation history. Further, when we correlate
the pixel spin chirality (not its absolute value but keeping the sign) with the star formation history we
find no correlation or to be precise the Pearson coefficient we find
is of order 4\% at all look-back times, which is below the error in the coefficient itself
(see below).

Fig.\ref{fig:chiralold} shows the relation between the pixel averaged star formation fraction ($SF_{j,1}$) in the oldest MOPED bin (1, see Table~1)  and the absolute value of the pixel-averaged spin $|S_j|$ for all those galaxies that have formed more than 30\% of their present stellar mass at $z > 2.0$. Because of the large number of pixels a scatter plot is unclear; here the points indicate the mean relation 
 and  the error bars show the standard deviation around the relation. Note that spiral galaxies that formed more than 30\% of their stars at $z > 2$ have preferentially a coherent spin chirality. If we focus on those spiral galaxies that formed more than 80\%, denoted by the dashed line, of their stars at $z > 2$ then virtually {\em all} of them have coherent spin chirality. This implies that those galaxies that were in place at $z > 2$ rotate on the same direction on scales of $\sim 10$Mpc.

To quantify our results we find that the Pearson correlation coefficient between
absolute  value of the pixel spin chirality and past star formation maps is maximum at
zero-lag for a  pixel physical size of $\sim 10$ Mpc/$h$.  For
smaller pixel sizes ($< 2$Mpc$/h$) the number of galaxies per pixel decreases very
rapidly. The correlation disappears for pixel sizes larger than $\sim
50$ Mpc/$h$. See also Fig.~1 where we show the value of the correlation for different pixel sizes.

We find that the correlation is greatest for galaxies which formed most
of their stars in the past and decays for galaxies with most of their
stars being formed recently.

Fig.~\ref{fig:corrspins} shows the Pearson correlation coefficient for
the absolute value of  the pixel spin chirality as a function of the star
formation bin (see Table~\ref{table:bins}). The points have been
obtained by computing the mean for several redshift slices of width
$\sim 100$ Mpc and the error bars shown are from the mean dispersion
in the different, $\sim 10$, redshift slices. The dashed line shows
the error in the correlation coefficient obtained from randomly redistributing the
 values of the $|S_j|$ map.  We can see that for those
galaxies with significant star formation in the oldest bins ($1-4$)
the correlation with absolute value of  pixel spin chirality is $6\sigma$
above the noise level, while it decreases to slightly weaker levels
for galaxies with significant star formation in their recent bins
(8-11). This finding indicates that spiral galaxies with most of their
star formation in the past have most of their galaxies rotating in the
same direction  with coherence  length of $10$ Mpc$/h$.  To explore the correlation
for separation smaller that $10$ Mpc/$h$ we have repeated the analysis
with the projected catalog. The catalog depth implies that in the line
of sight direction the cell size, $R$ is always $\sim 343$ Mpc/$h$
thus diluting the signal if the signal coherence length is smaller
than $R$. The higher density of objects however enables us to explore
smaller separations. We find that the signal is maximal for cell sizes
of $10$ Mpc/$h$ (given by the transversal cell size at $z_m$)
and decreases rapidly for cell sizes greater than $50$ Mpc/$h$. As before, also in this case when considering the full value of the pixel spin chirality, i.e. keeping the sign, we obtain no statistically significant correlation.

These results indicate  
that the universe does not have a preferred direction and/or more
importantly that human biases in classifying the rotation direction
do not enter into our analysis. 

\begin{table}
\begin{center}
\caption{Per cent deviation in each look-back time star bin for the quantity $|S_j|$ from zero, which means the universe has no preferred direction. Note that there is a negative bias at the \% level but this error is below the error from the correlation analysis (see text).}
\begin{tabular}{|cccccc|}
\hline
SF bin& 11 & 10 & 9 & 8 & $7 -1$ \\
\% & -0.7 & -2.5 & -3.4 & -2.9 & -3.8 \\ 
\hline
\end{tabular}
\end{center}
\label{table:spin}
\end{table}

Table~2 shows for the 11 
look-back times, the percentage deviation from zero in the value of the pixel spin chirality
when we consider the sign and  demonstrates that a small bias at the $3-4$\% 
level remains, toward negative values of $S$ but below the intrinsic error. To calculate the values of the bias in each look-back time bin we calculate $\sum S_{\beta}$ for all the galaxies in that bin that 
contribute to more than 90\% of the bin star formation.
We have also  verified that the strength of the  correlation signal does not depend on how many galaxies are included in the sample by randomly sub-sampling the catalog, the error-bars of course increase when less galaxies are considered.

Because star formation is also correlated with mass \citep{marks}, we
 explore if the above correlation is driven by the mass of the
galaxy. MOPED provides the present and total stellar mass in the
galaxy. In Fig.~\ref{fig:corrmass} we show the above correlation but
for three mass ranges.   The dashed line shows the correlation level for
a map where the absolute values of  pixel spins chirality have been redistributed randomly
and therefore is a measure of the noise level. We note that the
correlation at early times (bins 1-2) is dominated by galaxies in the
mass range $10^{10} - 10^{11}$ M$_{\odot}$. (Recall that since the sample is approximately volume limited the number of galaxies in each for the mass bins decreases rapidly with increasing  mass, so the trend is not driven by the number of galaxies in each mass bin). This is in contrast with
the strong correlation found between mass and star formation in the
past by \citet{Heavens,Panter07}. We conclude then that the observed
correlation between spin chirality and early star formation is not
driven by mass but by star formation activity.

So far we have used the galaxies from Galaxy Zoo classified as clean which  limits our 
sample to only $12897$ galaxies. Because there are many more galaxies in the MOPED catalog we 
have implemented the following algorithm to increase the Galaxy Zoo sample. We  weight the chirality of a galaxy by  the number of votes from the public. If most people voted for one direction this algorithm converges to the clean sample we used before. The results of the correlation so obtained are similar to Fig.~1 within the errors bars.

\section{discussion and conclusions}

Halo spin directions correlate with the large-scale structure see
e.g. \citet{BailinSteinmetz05,HahnSpins07,trujillo}: halos in sheets
tend to have their angular momentum parallel to the sheet, a weaker
indication is observed for halos in filaments to have their spin
perpendicular to the filament direction.

Spins directions are also correlated, as seen in observations
\citep{SlosarSpins, Pen2000} and as predicted by tidal torque theory
\citep{CatelanPorciani01,Catelanlensing,Crittenden2001,Porciani02a}.

We have found significant correlation between past star formation
activity and the degree of coherence of spin chirality for galaxies in a spatial patch (pixel spin chirality). While we have only
used the chiral information of nearly face-on galaxies, the detection
of a non-zero signal can only be produced if spin orientations are
correlated. The correlation is highest for regions in which galaxies have formed
most of their stars in the past.  We do not find significant correlations 
with spin coherence for any of the other measured galaxy properties that
we have considered (metalicity, stellar mass, metalicity history).
 
We note that using chiral information rather than axis inferred from
the projected ellipsoid is safer from the perspective of potential
systematic effects. The reason for this is that inferred properties of
stellar population are likely to be affected by the galaxy being
viewed face-on or edge-on. These properties, on the other hand, cannot
be affected by the galactic arms winding on way or another.

Our analysis indicates that neighboring spiral galaxies which have
similar star formation histories also have their spins aligned.  We can
interpret this as the large-scale environement influencing dark matter
halo spins, giving it a large scale coherence length, and that either
halo spin, environment, or a combination of the two, influence galaxy star formation
histories. This can be understood in the context of tidal-torque
theory.  Galaxies that have their spins aligned are formed in the same
filament or sheet and at about the same time.

This correlation is stronger for regions comprising galaxies with older stars, while
regions containing significant fractions of galaxies with recent star formation exhibit a smaller
correlation with spin coherence. This can be understood as these
galaxies being formed in the field and thus being affected by a random
tidal field and at different epochs or having had a recent major
merger, which would promote recent star formation but deviate the spin
direction from that set by the cosmological tidal field.

Our results imply that in simulations of galaxy formation one should expect to 
detect the early formation of big spiral galaxies in the filaments around clusters. It is the 
halos in the filaments that should carry the majority of the star formation of the big spirals. Further,
one would expect that most star formation today should be in the field, away from the filaments and 
that these objects should have randomly oriented spins.  
This finding indicate that spin and the correlation between spin and star formation are new, measurable  quantities which offer a new complementary way to explore and quantify the effect of environment on star formation \cite{Bamford}.
 We are planning to explore these issues using
cosmological simulations of galaxy formation. 

Another possible implication for these results involve weak
gravitational lensing studies. A weak lensing survey's potential to
yield a faithful reconstruction of the cosmological distribution of
dark matter is limited by the unknown intrinsic alignment of galaxy
shapes.  There is evidence that intrinsic galaxy shapes are spatially
correlated \cite{Brownetal2002,Pen2000,HeavensRefregierHeymans00}, due
to the fact that the alignment of galaxy disks orientations is induced
by halo spins correlations \cite{Catelanlensing}.

Our results seems to indicate that, if the star formation history
could be measured for at least some of the lensed galaxies, the
properties of their star formation histories could be used to predict
galaxies spin alignment and thus intrinsic shape alignment.
 
\section*{acknowledgments}
RJ and LV  are supported by FP7-PEOPLE-2007-4-3 IRG, FP7-PEOPLE-2007-4-3-IRG n. 202182 and by  MICINN (Spanish Ministry for Science and Innovation) grant AYA2008-03531. RJ and LV thank ICC at UB for hospitality.
RJ, AS, and LV hank the Galileo Galilei Institute for theoretical physics (GGI)  in Florence, where part of this work was carried
out, and INFN for partial support.

\end{document}